\documentstyle[aas2pp4,tighten,epsfig]{article}

\newcommand\degr{$^\circ$}
\newcommand\msun{M$_\odot$}

\newcommand{\ha}{H$\alpha$}
\newcommand{\hi}{H{\sc i}}
\newcommand{\hii}{H{\sc ii}}

\begin{document}

\title{Near infrared and optical morphology of the dusty galaxy NGC\,972}

\author{Y.D. Mayya$^1$, Swara Ravindranath$^2$ \and L. Carrasco$^{1,3}$}

\affil{$^1$Instituto Nacional de Astrofisica Optica y Electronica,  
           Apdo Postal 51 y 216, 72000 Puebla, Pue., M\'EXICO   }   
\affil{$^2$Indian Institute of Astrophysics, Koramangala, Bangalore 34, INDIA}
\affil{$^3$Also at Observatorio Astron\'omico Nacional, UNAM, M\'EXICO}

\affil{Electronic mail: ydm@inaoep.mx, cathy@iiap.ernet.in, carrasco@inaoep.mx}

\vskip1cm

\affil{To appear in {\it Astronomical Journal} October 1998}

\begin{abstract}

Near infrared (NIR) and optical surface photometric analyses of the dusty 
galaxy NGC\,972  are presented.
The photometric profiles in the $BVRJHK$ bands 
can be fitted with a combination of 
gaussian and exponential profiles, corresponding to a starburst nucleus
and a stellar disk respectively. The exponential scale length in the
$B$-band is 2.8 times larger than in the $K$-band, which implies a central
$B$-band optical depth as high as 11. 
A bulge is absent even in the NIR bands and hence the galaxy must
be of a morphological type later than the usually adopted Sb type. 
Relatively low rotational velocity and high gas content  
also favor a later type, probably Sd, for the galaxy.
Only one arm can be traced in the distribution of old stars; the second arm, 
however, can be traced in the distribution of dust and \hii\ regions. 
Data suggest a short NIR bar, which ends inside the nuclear ring.
The slowly rising nature of the rotation curve rules out a resonance
origin of the the nuclear ring. The ring is most likely not in the plane of 
the galaxy, given its circular appearance in spite of the moderately high 
inclination of the galaxy. 
The off-planar nature of the star forming ring, the unusually 
high fraction (30\%) of the total mass in molecular form, the 
presence of a nuclear starburst and the asymmetry of spiral arms, are 
probably the result of a merger with a gas-rich companion galaxy.

\end{abstract}

\section {Introduction}    

NGC\,972 is a nearby dusty galaxy whose morphological type is yet to be 
established.  It is classified as Sb in 
Hubble Atlas of Galaxies (Sandage 1961), Third Reference Catalogue of Bright
Galaxies (de Vaucouleurs et al. 1991, RC3 henceforth) and Revised 
Shapely-Ames Catalogue (Sandage \& Tammann 1981).
However, Krienke \& Hodge (1974) had found the galaxy to be sharing
many properties of I0 type galaxies and accordingly the galaxy 
was given an I0 classification in the Second Reference Catalogue of Bright 
Galaxies (de Vaucouleurs, de Vaucouleurs \& Corwin 1976). 
Given the heterogeneity of the I0 class galaxies, the 
presence of a well established trailing dusty arm (Burbidge, 
Burbidge \& Prendergast 1965) in this galaxy was taken as a 
strong support in favor of the Sb classification.
The presence of a star forming nuclear ring in this galaxy, if assumed 
to be a resonance ring, also seems to 
support the Sb classification of the galaxy 
(Ravindranath \& Prabhu 1998, RP98 henceforth). 

The optical appearance of NGC\,972, which was the basis for its
morphological classification hitherto, is dominated by the dust lanes. 
Interestingly the presence of these dust lanes has played a major role 
in its classification both as Sb and I0 --- the dusty trailing arm
supporting its Sb classification, the chaotic appearance caused by dust 
placing it in the class of I0 galaxies.
This highlights the limitation of the qualitative classification schemes based 
solely on the optical morphologies, especially for the dust-rich galaxies such 
as NGC\,972. Recent advances in observational techniques and theoretical 
modeling of galaxy dynamics allow us to use many more properties
than just optical morphology to establish the true morphological type 
of galaxies.  From the observational side, NIR imaging allows us 
to have a dust-free view of 
the underlying morphological components of galaxies. 
Such observations have already helped discover new components such as nuclear
spirals and bars in galaxies, whose optical appearance did not suggest 
the existence of those components (Zaritsky, Rix \& Rieke 1993; 
Knapen et al. 1995). 
The variation of dynamical properties, such as the rotation speed, length of 
the bar etc., along the Hubble sequence is now well understood. 
Early-type galaxies rotate faster than the late-type galaxies (Zaritsky 1993).
The bars of early-type galaxies are lengthy and uniform, where as
they are short and weak in late-type galaxies (Combes \& Elmegreen 1993). 
Thus galaxy  classification can be done in an integrated way
by making use of both the morphological and dynamical properties of galaxies. 
Such studies will also help in isolating the underlying physical quantities
governing the Hubble sequence.

We carried out a quantitative morphological investigation of NGC\,972 to 
check whether its Sb classification requires revision. We based our 
analysis on the newly
obtained NIR images, and made use of the existing optical images. 
Additionally we compiled the available information on the gas content, 
star formation indicators and 
dynamics to infer its morphological type in an integrated way.
We found that none of the quantitative indicators of
the galaxy morphology favor its Sb classification and the galaxy seems to be
of type as late as Sd.
Observations and the techniques followed in the reduction of the data
are described in Sec.~2. The main surface photometric results are
presented in Sec.~3 and the opacity of the disk of NGC\,972 is discussed in
Sec.~4. Issues related to the morphological type and the dynamical history
of the galaxy are discussed in Sec.~5. Concluding remarks are given Sec.~6.
A distance of 21.9~Mpc is adopted for the galaxy, based on the velocity 
from RC3 and Hubble constant of 75 km\,s$^{-1}$\,Mpc$^{-1}$.
This results in an image scale of 105 parsec arcsec$^{-1}$.

\section{Observations and Data Reduction}   

Near infrared observations were carried out using the CAMILA instrument
(Cruz-Gonzalez et al. 1994)
at the 2.1-m telescope of the {\it Observatorio Astron\'omico Nacional} 
at San Pedro Martir, Baja California. The CAMILA instrument uses a NICMOS 3 
detector of 256$\times$256 pixel format. The instrument was used in the imaging 
mode with the focal reducer configuration $f/4.5$  
in all our observations. This results in a spatial resolution of 
 $0\farcs85/$pixel and a total field of view of $3\farcm6\times3\farcm6$. 

The imaging observations were carried out on the night of 13 September 1997
using the broad band filters $J,H,K'$. Each observation consisted of a sequence
of object and sky exposures, with the integration time of an individual 
exposure limited by the sky counts, which was kept well below the 
non-linear regime of the detector. The net exposure times 
on the object were 12, 6 and 5 minutes for $J,H$ and $K'$ bands respectively.
Roughly equal amounts of time were spent on the sky fields.
The recessional velocity of 1670~km\,s$^{-1}$ for NGC\,972 did not allow us
to detect the Br$\gamma$ emission line with the available 
zero-redshift Br$\gamma$ filter.
Photometric calibration to the standard $JHK$ system was performed
using the UKIRT standard star FS\,7 (Casali \& Hawarden 1992). 
The sky conditions were photometric and the sky magnitudes were roughly
$15.9, 14.2$ and 11.5 magnitude arcsec$^{-2}$ in $J,H,$ and $K$ bands 
respectively. The seeing FWHM was typically $2\farcs0$. 

The image processing involved subtraction of the bias and sky frames,  
division by flat field frames, registration of the images to a common 
co-ordinate system and then stacking all the images in a filter. Bias 
subtraction was carried out as part of the data acquisition. The sky frames 
in each filter were prepared by combining all the sky frames using the 
median operation. Flat field frames were taken during twilight period. 
A master flat field frame for each filter was constructed by stacking
several night-sky subtracted twilight frames in corresponding filters. 
The flat fielding operation involved dividing the sky subtracted images of the 
object by normalized master flats.
The resulting images were aligned to a common co-ordinate system using the 
common stars in the frames and then combined using the median operation. 
Only good images (as defined in the CAMILA manual --- 
see Cruz-Gonzalez et al. 1994) were used for combining the images. 
Some of the images had a horizontal band at the joints of the 
individual chips forming the detector of the NICMOS 3 camera. We eliminated 
this feature by subtracting a $255\times1$ median smoothed image from the 
original images.
The resulting combined images were aligned to the image from the 
Digitized Sky Survey (DSS), through a geometrical mapping using the GEOTRAN
and GEOMAP tasks available in the reduction software. The transformed star 
positions in the images agreed to within $0\farcs2$ as judged from the 
coordinates of common stars.
Also, the images were assigned the equatorial coordinates using the image
from the Digitized Sky Survey. 

The routines under Image Reduction and Analysis Facility (IRAF\footnote 
{IRAF is distributed by National Optical Astronomy Observatories, which are 
operated by the Association of Universities for Research in Astronomy, Inc., 
under cooperative agreement with the National Science Foundation.}) and
Space Telescope Science Data Analysis System (STSDAS) were used
in the reduction and analysis of all the data. Optical images used in
this study are from RP98. 

\section{Morphological Structures}   

The most striking feature on the optical images of the galaxy is a dust lane 
running from southeast to northwest. The prominent spiral arm
is on the northwest, partially broken by the dust lane. The second
spiral arm in the symmetric position on the southeast is traced 
by the dust lane. The appearance of the dusty arm suggests that the 
southeast side is the near side of the galaxy. Burbidge, Burbidge \& 
Prendergast (1965) used this information along with the rotation curve 
to establish that the spiral arms are trailing in this galaxy. 
However, the stellar arm corresponding to the dusty arm cannot be traced 
on the optical images which is probably due to the heavy obscuration.

The structure of the galaxy in the optical and NIR bands is shown as
grey scale maps in Fig.~1. The $K$ and $B$ band images are shown 
in Figs~1(a) and (b) where as the $H-K$ and $J-K$ color maps are
shown in Figs~1(c) and (d) respectively.
The galaxy has a smooth distribution of intensity in the NIR bands as
compared to the optical bands. The dust lanes, which are prominent in 
the $B$ band image are absent in the $K$ band image. The northwest spiral arm
is easily traceable and continuous in all the three NIR bands and
coincides in position with that in the $B$-band. However the southeast stellar
arm, which is expected to lie on the leading side of the dusty arm,
cannot be traced even on NIR images. 
Instead, a linear spur can be traced on the NIR images on the inner side of the
dust lane. The spur lies roughly parallel to the major axis of the galaxy
and runs $\sim5\arcsec$ south of the nucleus into the northwest side.
The nucleus and the optically prominent \hii\ region on the northwest can 
both be traced in 
the NIR bands, with the nucleus relatively brighter than the \hii\ region. 
A bulge is not apparent on the NIR images. A short bar-like
extension along the northsouth direction can be seen.

\subsection{Spiral arms, Dust lanes and Star formation}   

Images in the $K$-band are least affected by dust and hence are best
suited to study the intrinsic components, such as bulge, disk and the 
spiral arms of galaxies. 
On the other-hand the $B$-band images are heavily affected by the dust 
and hence structural differences between $B$ and $K$ bands can be used to map
the distribution of dust. Difference in stellar populations can also lead 
to structural differences between $B$ and $K$ bands (de Jong 1996). 
However in dusty galaxies such as NGC\,972, the major contribution to
the observed structures on color images comes from the non-uniform
distribution of dust, rather than the stellar population gradients.
Thus it can be assumed that the structures seen in $B-K$ color
image of NGC\,972, to first order, represent the dust distribution. 
Contours of $B-K$ colors, drawn to illustrate the dust lanes, are shown 
superimposed on the $H-K$ and $J-K$ color maps (Figs~1(c) and (d)).
The outer contour almost forms the boundaries of the structures seen in 
the NIR color maps, indicating that the dust absorption is non-negligible 
even at NIR wavelengths in this galaxy.

Distribution of giant \hii\ regions, as traced by H$\alpha$ emission,
characterizes the regions of current star formation. 
\ha\  emission in NGC\,972 extends to 3.4~kpc (RP98) and is shown as 
a grey scale map in Fig.~2.
$K$-band contours at levels selected to illustrate the disk shape,
spiral arms and the nuclear regions are superimposed (thin double lines 
denoted by 1, 2 and 3 respectively)
on the \ha\ map. The boundary of the principal structures seen on 
the $B-K$ color map are shown as contours (thick lines) in this figure.
On the southeast side, these contours demarcate the leading edge of a 
dusty arm.

It can be seen that the star formation extends almost over the entire stellar
disk as traced by the $K$ band isophote. 
Current star formation is active along the northwest 
spiral arm. \hii\ region population can be traced along
the southeast dusty arm as well. Significantly, these \hii\ regions lie
on the convex side of the dusty arm, as is expected for a trailing 
density wave. \hii\ regions near the center of the galaxy are distributed
along a partial nuclear ring of radius $6\arcsec$. Note that this ring is 
nearly circular in shape,
whereas the outermost $K$-band isophote (contour numbered as 1) is 
highly elliptical. The central $K$-band contour (denoted as 3) is of
oval shape and is elongated along the line connecting the beginning of
the spiral arms. This suggests the presence of a weak bar in the galaxy, 
which seems to end just inside the circumnuclear ring. There is significant 
amount of star formation in the inter-arm regions of this galaxy. 

\subsection{Bar, Bulge, Nucleus and Disk}   

Azimuthally averaged radial surface brightness profiles have been most
widely used for a quantitative analysis of the morphological structures.
$K$-band profiles are best suited for this purpose because of their ability 
to trace the old stellar population and also because they are least 
affected by the obscuring dust. However the high background 
levels in the $K$-band restrict the usable radius to within $1\arcmin$
of the center. On the other hand $J$ band profiles are
less affected by the sky background and hence we carried out surface
photometric analysis on the $J$ profiles as well.
Surface photometric analysis included fitting ellipses to the isophotes and
the construction of the radial profiles of isophote intensity, 
ellipticity ($\epsilon$) 
and the position angle (PA) of the resulting ellipses. These were
carried out using the ellipse fitting routines under the STSDAS package. 

The results of surface photometric analyses are plotted in Fig.~3. 
The ellipticity and PA for the $J$ and $K$ bands are plotted in Figs.~3(a) 
and (b) respectively. Both the plotted quantities attain a steady value
(ellipticity $=0.60$ and PA$=$152\degr) at radii exceeding $35\arcsec$.
The resulting PA is exactly equal to the optically measured
value (RC3). However the value of ellipticity suggests that the galaxy is 
marginally more elongated in the NIR compared to the optical value in RC3. 
Figures such as 3(a) and 3(b) can be used to infer the presence of a bar;
e.g. a strong bar would show up as a high eccentricity and a 
constant position angle structure between the nucleus and the beginning 
of the spiral arm. No such structure is obvious in Figs~3(a) and (b), 
and hence a strong bar is absent in this galaxy. However there is an
indication for the presence of a small bar at PA$=170$\degr,
which ends inside a radius of 10$\arcsec$. This weak bar can
also be seen on the $K$-band image (Fig.~1(a)) as discussed earlier.

A quantitative estimate of the bulge-to-disk ratio of galaxies can be
done by decomposing the radial intensity profiles into de Vaucouleurs
$r^{\onequarter}$ and exponential profiles. 
We obtained azimuthally averaged radial intensity profiles, by fixing
the eccentricity and PA corresponding to their
values at the outer radii.
We fitted the radial intensity profiles with a composite 
profile containing a nucleus (gaussian), bulge ($r^{\onequarter}$) 
and an exponential disk. The profile shape is inconsistent with the 
presence of an $r^{\onequarter}$ central bulge for both $J$ and $K$ profiles. 
In Fig.~3(c), we show the observed intensity profiles
in $J$ and $K$ bands along with the best fit combination of gaussian
and exponential profiles.
The central component corresponds to the nucleus with a gaussian intensity 
profile of $\sigma=4\farcs0$ (420~pc). The fitted exponential disk has  
scale lengths of 1.2 and 1.5~kpc in $K$ and $J$ bands respectively.

Thus we conclude that there is no bulge in NGC\,972. The central 
bright spot corresponds to the nucleus. 
The bar is weak and ends inside a radius of 1~kpc. 
The disk intensity follows an exponential profile, with a scale length
of 1.2~kpc in the $K$-band.

\subsection{Surface brightness and color profiles}   

Azimuthally averaged radial intensity profiles in $BVRJHK$ bands and in
the emission line of \ha\ were obtained by fixing the eccentricity and 
PA corresponding to their values at the outer radii.
Prior to profile extraction, 
the point spread functions of all the images are matched (using the IRAF
task PSFMATCH) to the one having the poorest seeing ($2\farcs7$).
The resulting intensity and color profiles are shown in Figs~3(d) and (e) 
respectively. It can be easily noticed that the intensity profile in broad 
bands is the steepest in the $K$-band and the flattest in the $B$-band. 
The profiles flatten systematically with decreasing wavelength of the bands. 
This results in a color gradient with the disk becoming gradually 
bluer at outer radii. The nuclear $B-K$ color reaches as red as 5.5.
The $B-K$ color profile shows a red bump between 20\arcsec\ and 30\arcsec\ 
which corresponds to the position of the dusty arm.

The \ha\ emission line surface brightness, which is a measure of 
current star formation, is expressed in magnitude
units (with an arbitrary zeropoint) and hence its profile shape
can be directly compared with those in the broad bands. 
It distinctly differs from other profiles between 10\arcsec and 25\arcsec, 
where the profile falls slower than the brightness profiles of old disk stars. 
This is due to the presence of the brightest \hii\ region in the galaxy
in this zone. Outside this zone \ha\ surface brightness falls smoothly.

\section{Scale lengths and Central Optical Depth}   

The question about the opacity of galactic disks has drawn a lot of
attention in recent years (Burstein, Haynes, \& Faber 1991). 
The availability of data extending from $B$ to $K$ bands allows us to 
study the opacity of the disk in NGC\,972.
One of the quantities which is extensively used for this purpose is the
scale length of the exponential disks in different bands 
(Bothun \& Rogers 1992; Evans 1994; Peletier et al. 1994).
The underlying stellar disks of galaxies do not show a strong color gradient 
and hence the intrinsic (dust-free) scale length of a galactic disk is only 
weakly dependent on wavelength. However, absorption by dust has the effect of 
flattening the observed intensity profiles, the effect being maximum at 
shorter wavelengths. Thus for dusty disks, the scale length observed
at $B$ band is expected to be larger than that at the $K$ band.

Evans (1994) modeled the wavelength dependence of scale lengths of galaxies 
for dusty disks. He found that the amount of increase in the scale length 
at shorter wavelengths depends on the amount of dust (or optical depth) 
as well as on the relative scale height of dust with respect to that of 
stars in disks of galaxies. While comparing the observed scale lengths 
available in the literature to his model results, he noted the necessity
to use the same radial zones of intensity profiles to obtain the 
scale lengths at different wavelengths. Following his suggestions, we obtained 
scale lengths in $BVRJHK$ bands by fitting exponential profiles between
an inner radius of 25\arcsec\ and an outer radius of 65\arcsec\ for all the 
profiles. The resulting scale length in the $B$-band 
is r$_{\rm d}(B)=28.6\arcsec=0.3R_{25}=3$~kpc. Ratios of $B$-band
scale length to that in other bands is given in Table 1.
Evans (1994) computed scale lengths at $B, I$ and $H$ bands for galaxies
with and without bulges. Among the computed pure disk models, a model
with dust scale height equal to that of stars can produce the observed
scale length ratio r$_{\rm d}(B)/\rm{r}_{\rm d}(H)=2.1$
for a central optical depth $\tau_B(0)=11$. 
Models including a bulge imply even higher $\tau_B(0)$ value.
Moriondo, Giovanardi, \& Hunt (1998) had obtained a mean value 
for the central optical depth 
in $V$-band $\tau_V(0)=$2--4, for a sample of early-type spiral galaxies. 
Hence the inferred value of optical depth in NGC\,972 is higher than
for normal galaxies, which is not surprising given its dusty appearance. 
Such high values are also inferred in other dusty galaxies; 
for example, Evans (1994) had estimated a value
of $\tau_B(0)=20$ in the dusty starburst galaxy NGC\,253.
 
\begin{deluxetable}{ccc}
\tablenum{1}
\tablewidth{0pt}
\tablecaption{Scale lengths of NGC\,972\tablenotemark{1}}
\tablehead{
\colhead{Band}   &  \colhead{r$_{\rm d}(\lambda)(\arcsec)$} &  
\colhead{r$_{\rm d}(B)$}/{r$_{\rm d}(\lambda)$}  \\  
}
\startdata
$K$    &  10.23  &  2.76 \\
$H$    &  13.46  &  2.10 \\
$J$    &  16.41  &  1.72 \\
$R$    &  23.42  &  1.20 \\
$V$    &  24.39  &  1.16 \\
$B$    &  28.21  &  1.00 \\
\enddata
\tablenotetext{1}
{Disk profiles between 25\arcsec\ and 65\arcsec\ are used in obtaining
the scale lengths.}
\end{deluxetable}

\begin{deluxetable}{llll}
\tablenum{2}
\tablecaption{Morphological and Starburst properties of NGC\,972}
\tablehead{
\colhead{Property}   & \colhead{Value} & \colhead{Comment} & 
\colhead{Reference to Col.~2} \\  
}
\startdata
Bulge/Disk ratio             & no bulge  &     Sd or later & This work\\
Pitch Angle ($^\circ$)       & 50--60    &     Sd or later & This work\\
V$_{\rm rot}^{\rm max}$ 
(km s$^{-1}$)                & 120       &     Sc or later & Burbidge et al. (1965)\\
Mass ($10^{10}$ M$\odot$)    & 1.2       &     Sc or later & Burbidge et al. (1965) \\
M(HI)/M(T)                   & 0.21      &     Sd or later & Young et al. (1996)\\
EW(\ha) (\AA)                & 36.4      &     Sc or later & RP98 \\
$B-V$                        & 0.64      &     Sb or later & RC3 \\
$U-B$                        & 0.07      &     Sb or later & RC3 \\
Yerkes Type                  & F3        &     $\sim$ Sc & Humason et al. (1956)\\
     & & \\
Mean Type                    & \nodata   &     Sd  & \\
     & & \\
L(FIR) ($10^{10}$ L$\odot$) & 3.67 & $1.15\times$M\,82 & Young et al. (1996) \\
L$_{{\rm H}\alpha}$ 
($10^{41}$ erg s$^{-1}$)      & 2.95 & $0.85\times$M\,82 & RP98, Young et al. (1996) \\
SFR 
(M$\odot$ yr$^{-1}$)& 2.64 & $0.85\times$M\,82  & RP98 \\
M(H2)/M(HI)         & 1.41& $1.70\times$M\,82 & Young et al. (1996)\\
M(HI+H2)/M(T)       & 0.50& \nodata  & Young et al. (1996)\\

\enddata
\end{deluxetable}

\section{Discussion}   

\subsection{The nuclear \ha\ ring}   

RP98 reported the presence of a nearly circular star forming ring of 
radius 630 pc around the nucleus of NGC\,972. Given that the inclination 
of the galactic disk to the line of sight is 60\degr, the nuclear ring is 
either in the galactic plane and intrinsically elliptical or it is 
off-planar and intrinsically circular.
Classical nuclear rings are circular and lie in the plane of the parent
galaxy and are commonly found in early-type barred galaxies. 
Such rings are associated with the Inner Lindbland Resonance (ILR) of galaxies. 
The existence of ILR in a galaxy depends on the 
form of the rotation curve in the central parts of galaxies and the 
pattern speed of the bar or the spiral arm. Burbidge et al. (1965) 
had obtained the rotation curve for the galaxy from slit
spectroscopy at different position angles. They found that the rotation
curve in the galaxy rises very slowly reaching values of around 
100~km\,s$^{-1}$ at a radius of 25$\arcsec$. The galaxy rotates 
like a solid body at least up to 10$\arcsec$ (1 kpc). ILR can exist
only outside the solid body rotating region and hence the observed ring 
at 630~pc is not a resonance ring.

The \hii\ regions forming the ring can be traced in the optical broad bands, 
especially in the $B$-band, and cannot be traced in any of the NIR bands. 
In general nuclear rings in galaxies have stronger continuum compared to
the disk \hii\ regions (Kennicutt, Keel \& Blaha 1989). This is understood 
in terms 
of a longer history of star formation in circumnuclear \hii\ regions 
(Korchagin et al. 1995). The absence of a strong
continuum in the nuclear ring of NGC\,972 indicates that the star formation 
in the ring has started relatively recently. The formation of the ring
is possibly associated with the perturbation caused by a small
intruding galaxy (see Sec.~5.3).

\subsection{Revision of morphological type of NGC\,972}   

The absence of a bulge even on the NIR images calls for a re-discussion 
of the adopted Sb morphological classification of NGC\,972. 
In recent years, there have been reports of the absence of classical
bulges in galaxies classified as early-type spirals (Carollo et al. 1997). 
These may either represent pure mis-classifications, considering the 
subjective nature of the classification, or intrinsic limitations of the
criteria used in classification. For example, one of the intrinsic limitations 
was  recently demonstrated by Combes and Elmegreen (1993). Using 
numerical simulation, they found that bar morphology
is more tightly related to the dynamical properties rather than the
spiral morphology of the galaxy. Thus a bar with early-type properties
can be present in a late-type spiral, as was found recently in NGC\,6221 
by Vega-Beltr\'an et al. (1998). Historically used classification criteria do
not allow for such mixed characteristics in galaxies.

We compiled all the existing global properties on NGC\,972 in an attempt to 
clarify its morphological type. In particular, we aim to establish whether 
it is a mis-classified galaxy or has mixed morphological characteristics.
The compiled data are presented in Tab.~2. 
The parameters in the upper half of the table are related to the 
morphological type whereas those in the lower part 
are related to the starburst properties. The dependence of the 
listed parameters with morphological type is well established statistically 
(e.g. Roberts \& Haynes 1994).
The most likely morphological type for NGC\,972, based on
each of the observed quantities, is indicated in Col.~3. 
Principal morphological indicators, namely bulge-to-disk ratio and the 
pitch angle (of the north-west arm measured on the deprojected $K$-band
image), are clearly inconsistent with the Sb classification.
The dynamical mass, as inferred from the rotational speed, is also too small 
for the Sb classification. 
The size of the weak bar is almost equal to the exponential scale length --- a 
condition found to be typical in late type galaxies (Combes \& Elmegreen 1993). 
Secondary indicators of morphological type such as the neutral hydrogen 
content and the \ha\ equivalent width
(a measure of present to past star formation rate --- see Kennicutt \& Kent
1983), also suggest a morphological type later than Sc for the galaxy. 
Observed colors are redder than those found in galaxies of Hubble types 
Sc and later. 
However the prominent stellar population, as indicated by the Yerkes type, 
is not consistent with the red colors, suggesting that the galaxy may 
have intrinsically blue color typical of late-type spirals, 
but is reddened by the heavy amount of dust in the galaxy.
Thus all the global properties suggest a morphological type later than Sc.
The galaxy however has several peculiarities, the most striking one
being the absence of a stellar arm accompanying the dusty arm on the
southeast side. 
Thus the most appropriate morphological type for NGC\,972 would be SABd pec.

\subsection{Is NGC\,972 result of a minor merger?}   

The Far Infrared (FIR) and \ha\ luminosities suggest a high rate of recent 
star formation in NGC\,972. Nuclear starburst, whose properties compare 
well with that of other well-known starbursts, contributes to most of this 
star formation (Table 2; see also RP98). 
Starburst activity is common in galaxies having an interacting companion 
or in galaxies formed due to merging of two nearly equal mass systems.
The chaotic distribution of dust lanes in NGC\,972 
gives it a close resemblence in appearance to the starburst galaxy M\,82.
The present activity of M\,82 is known to be triggered due to its
interaction with M\,81 (Ichikawa et al. 1994; Yun, Ho \& Lo 1994). 
However there is no galaxy visible on the Palomar Sky 
Survey prints within 10 times the 
optical diameter with a velocity difference less than 1000 km\,s$^{-1}$
of NGC\,972 (Solomon \& Sage 1988). The galaxy does not show any obvious
signatures of a merger such as tidal tails or bridges. Thus it is safe to
assume NGC\,972 is an isolated galaxy. Among the isolated galaxies,
starbursts are most commonly associated with a strong bar. Absence of a 
strong bar in NGC\,972 calls for invoking alternative mechanisms for
triggering the activity in this galaxy. One of such alternative
mechanisms is a merger with a low-mass
companion object, such as a dwarf galaxy (minor merger).

NGC\,972 at present contains an unusually high amount of its mass (50\%) in 
the gaseous form, with the molecular mass exceeding the atomic mass 
(Young et al. 1996).
Such a high ratio of molecular to atomic mass is typical of that found 
in optically distorted and merging galaxies (Mirabel \& Sanders 1989), 
which gives independent support for a possibly merger-induced starburst 
activity in this galaxy. NGC\,972 group of galaxies contains many dwarf 
spheroidals (Vennick \& Richter 1994) and hence a merger with one of those 
galaxies in the recent past cannot be ruled out. 
If that is the case, the merged dwarf spheroidal galaxy had to be gas-rich. 
However dwarf spheroidal galaxies are normally poor in gas content.
On the other hand, small gas clouds (mass less than 10\% of the
primary galaxy) are known to exist around several galaxies 
(Schulman, Bregman \& Roberts 1994), the most familiar example being the
high velocity clouds around our own Milky Way.
Accretion of such clouds can naturally enhance the gas mass in the galaxy.
As the accreted gas flows to the center of the galaxy, it transforms to  
molecular form, triggering the intense burst seen in the galaxy. 
The accretion process is probably responsible for the off-planar nuclear ring. 
In such a scenario, the plane of the ring may represent the plane in 
which the gas is being accreted.

NGC\,972 is very asymmetric with its northwest arm much stronger than the 
southeast arm in the stellar continuum. Such an asymmetry is
common in late-type galaxies containing a companion (Odewahn 1994). 
In a recent study, Pisano, Wilcots \& Elmegreen (1998) argue that the 
observed morphological and kinematical asymmetries in the late type galaxy 
NGC\,925 are due to one or many interactions with a companion
low-mass galaxy. They discovered an \hi\ cloud of mass $10^7$~\msun\ in 
the neighborhood of the galaxy, which they believe is the 
residual gas cloud resulting from the interactions.
Thus it is very likely that
the observed asymmetry in NGC\,972 is caused by the minor merger
of a gas-rich companion, which as we discussed above, can also account
for the observed starburst activity, 
high molecular gas fraction and the off-planar nuclear ring. 
It however remains to be seen whether gas clouds, such as that found
in NGC\,925, also surround NGC\,972. This is one of the issues we will
be investigating in the future.

\section{Conclusions}  

We carried out a detailed analysis of the morphological type of NGC\,972
using a variety of physical parameters. We favor a morphological type as
late as Sd based on the absence of a bulge, high pitch angle of the
spiral arm, low dynamical mass and high mass fraction in gas.
However, it was classified as Sb and I0 in major astronomical catalogs 
and atlasses, which were mainly guided by the morphological appearence 
of dust lanes rather than a detailed quantitative analysis such as carried
out in our work.
The galaxy contains heavy amount of dust with central B-band face-on
optical depth as high as $\tau_{\rm B}(0)=11$. The spiral arms are
asymmetric with the northwest spiral arm brighter than the southeast 
spiral in the stellar continuum. The galactic nucleus is undergoing a 
starburst with a strength comparable to that in M\,82. In addition 
there is active star formation in an off-planar nuclear ring, and the
galaxy is extremely gas-rich, especially in molecular form. We propose
that all these activities are a result of merger of NGC\,972 with a gas-rich
companion.

\begin{acknowledgements}
We thank the time allocation committee of the {\it Observatorio 
Astron\'omico Nacional} for granting us the telescope time
for NIR observations. We also thank Olga Kuhn for her excellent support
as resident astronomer at the observatory and Ivanio Puerari for
reading the manuscript. The Digitized Sky Surveys were produced at the
Space Telescope Science Institute under US government grant NAGW-2166.
The images of these surveys are based on photographic data obtained using
the Oschin Schmidt Telescope on Palomar Mountain and the UK Schmidt 
Telescope.

\end{acknowledgements}

\clearpage


\begin{figure}[ht]
\vspace*{-2cm}
\hspace*{3.5cm}
\centerline{\psfig{figure=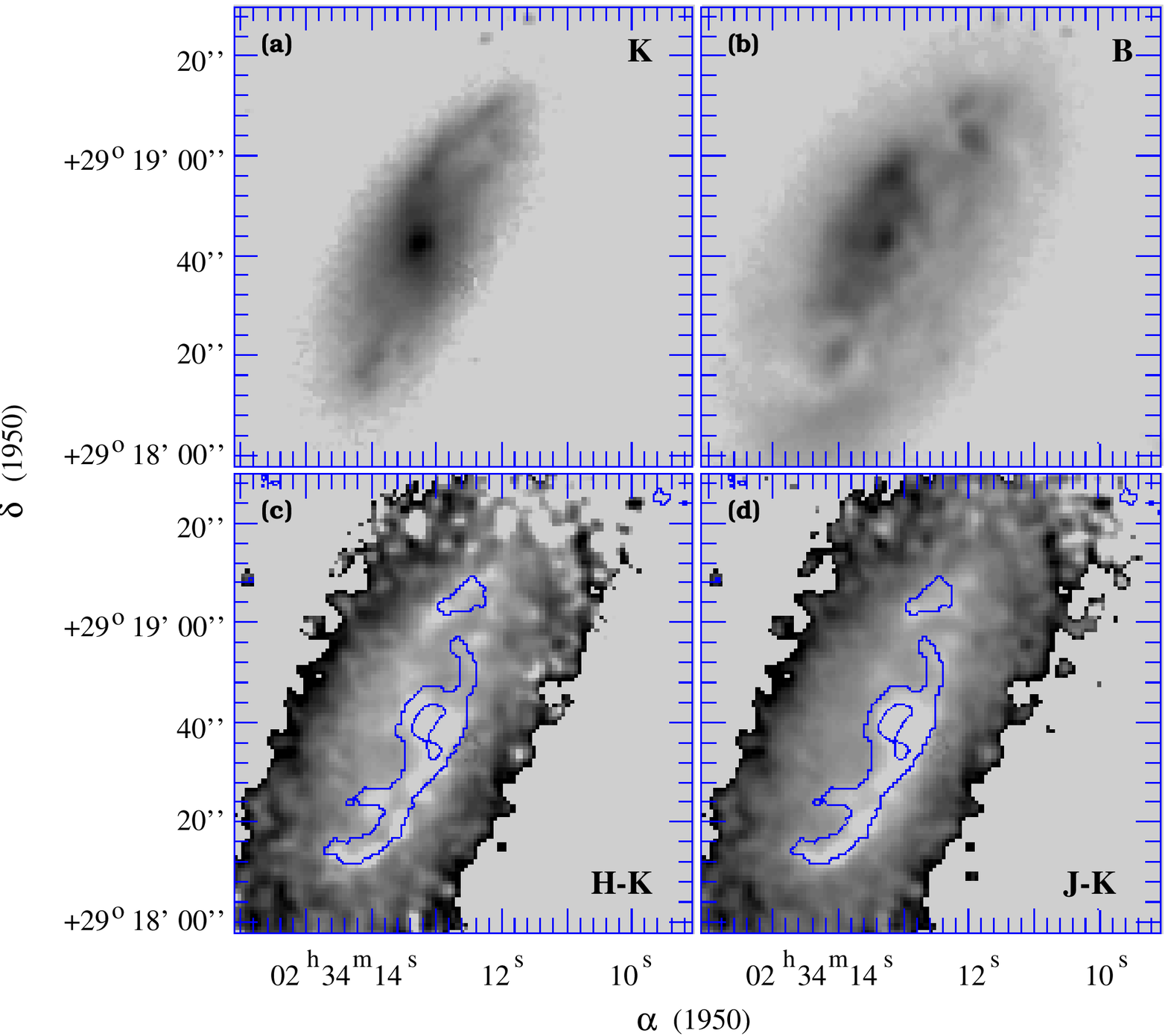,height=15cm}}
\vskip 1cm
\caption{Grey scale images of NGC972 in (a) $K$-band (b) $B$-band 
(c) $H-K$ color and (d) $J-K$ color (brighter grey scales correspond 
to redder colors). Contours of $B-K$ color are superimposed
on the color maps and represent the position of the optically seen dust lane.
}
\end{figure}

\clearpage

\begin{figure}[ht]
\vspace*{-2cm}
\hspace*{2cm}
\centerline{\psfig{figure=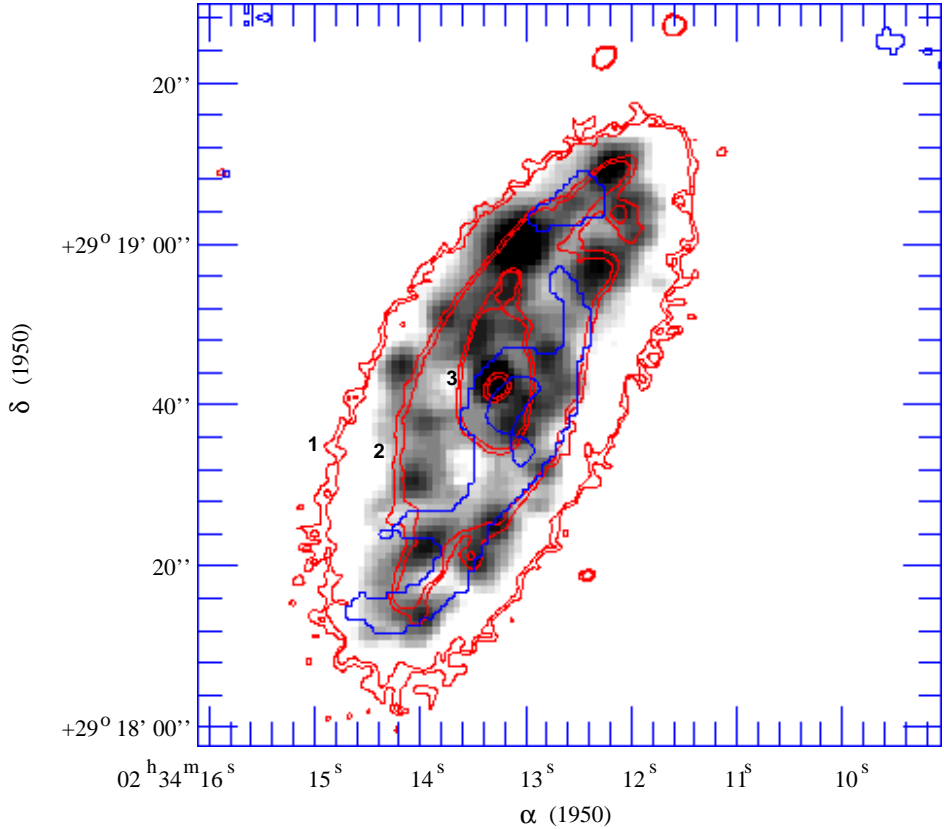,height=11cm}}
\vskip 1cm
\caption{Iso intensity contours in $K$ band (thin double line) and $B-K$ color 
(thick line) are superimposed on the grey-scale \ha\ image of NGC972.
Four pairs of $K$ contour levels are drawn and illustrate the outer structure,
the spiral arm, central weak bar (numbered 1, 2 and 3 respectively) and the 
nucleus. These contours correspond to surface brightness levels
18.4, 16.9, 15.7 and 14.2 magnitude arcsec$^{-2}$ respectively. 
Two $B-K$ contours are drawn, corresponding to colors 4.75 (outer) and 5.50
magnitude. The outer $B-K$ contour traces the dust lane, running all the
way from the south-east to the northwest spiral arm. Notice that the
distribution of disk \hii\ regions follows the $K$-band spiral in the northwest 
and  dust lane on the southeast side. Another notable feature is the
vastly different ellipticities of \ha\  nuclear ring (nearly-circular) 
and the outer $K$-band isophote.
}
\end{figure}

\clearpage

\begin{figure}[ht]
\vspace*{-1cm}
\hspace*{3.8cm}
\centerline{\psfig{figure=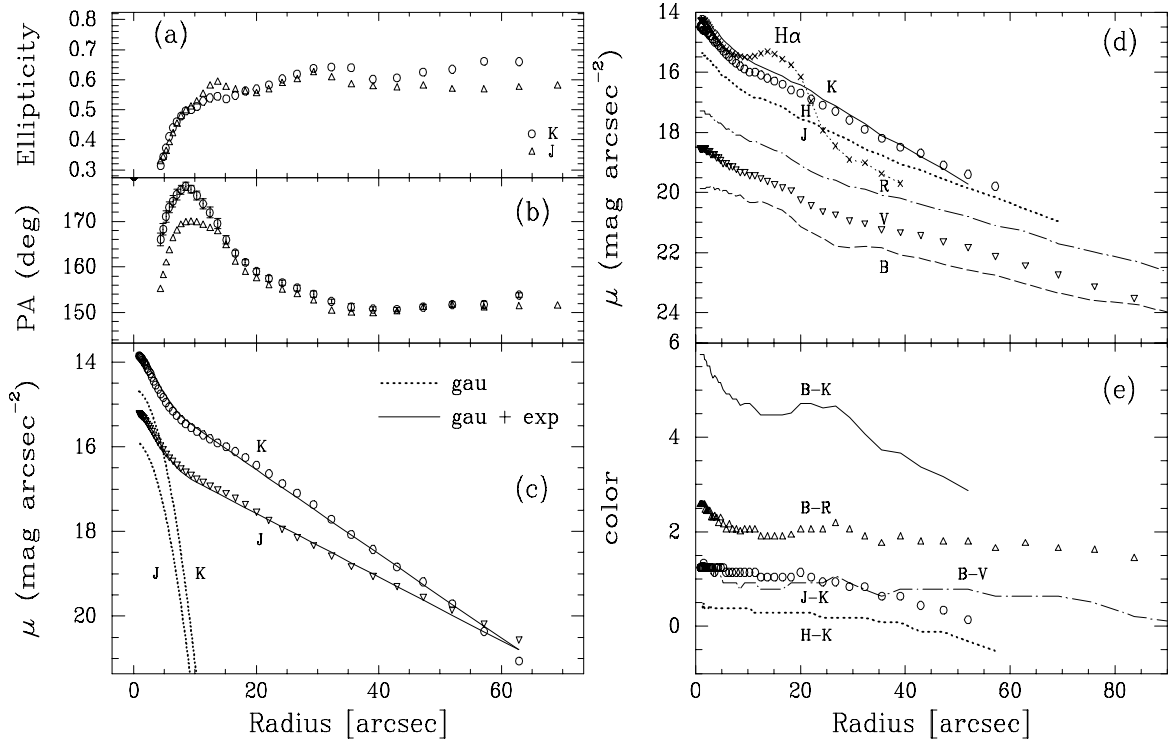,height=11cm}}
\vskip 1cm
\begin{center}
\caption{Surface photometric analysis of NGC972. 
Ellipticity (a), the major axis position angle (b) and the surface brightness 
(c) of the isophotes 
for $J$ and $K$ bands are plotted. The observed profiles can be well-fitted
by a combination of gaussian (nucleus) and exponential (disk) profiles. 
The surface brightness profiles in $BVRJHK$ bands and the corresponding 
color profiles are plotted in (d) and (e). For comparison \ha\   
surface brightness profile (expressed in magnitude units with an arbitrary 
zeropoint) is also plotted. Note the systematic
flattening of the surface brightness profiles from $K$ to $B$-band --- 
a definite signature of optically thick disks. 
}
\end{center}
\end{figure}

\end{document}